\documentclass[pre,aps]{revtex4}
\usepackage{graphicx}

\begin{document}

\title{Low-temperature thermodynamics of the classical frustrated
ferromagnetic chain in magnetic field}
\author{D.~V.~Dmitriev}
\email{dmitriev@deom.chph.ras.ru}
\author{V.~Ya.~Krivnov}
\author{N.~Yu.~Kuzminyh}
\affiliation{Joint Institute of Chemical Physics, RAS, Kosygin str. 4, 119334, Moscow,
Russia.}
\date{}

\begin{abstract}
Low-temperature magnetization curves of the classical frustrated
ferromagnetic chain in the external magnetic field near the transition point
between the ferromagnetic and the helical phases is studied. It is shown
that the calculation of the partition function in the scaling limit reduces
to the solution of the Schr\"{o}dinger equation of the special form for the
quantum particle. It is proposed that the magnetization of the classical
model in the ferromagnetic part of the phase diagram including the
transition point defines the universal scaling function which is valid for
quantum model as well. Explicit analytical formulae for the magnetization
are given in the limiting cases of low and high magnetic fields. The
influence of the easy-axis anisotropy on the magnetic properties of the
model is studied. It is shown that even small anisotropy essentially changes
the behavior of the susceptibility in the vicinity of the transition point.
\end{abstract}

\maketitle

\section{Introduction}

Strongly frustrated low-dimensional magnets have attracted much attention
last years \cite{mikeskabook}. A very interesting class of such compounds is
edge-sharing chains where $CuO_{4}$ plaquets are coupled by their edges \cite%
{Mizuno,Masuda,Hase,Malek,Capogna,Nitzsche}. An important feature of the
edge-sharing chains is that the nearest-neighbor (NN) interaction $J_{1}$
between $Cu$ spins is ferromagnetic while the next-nearest-neighbor (NNN)
interaction $J_{2}$ is antiferromagnetic. The competition between them leads
to the frustration. A minimal model describing the magnetic properties of
these cuprates is so-called F-AF spin chain model the Hamiltonian of which
has the form
\begin{equation}
H=J_{1}\sum \mathbf{S}_{n}\cdot \mathbf{S}_{n+1}+J_{2}\sum \mathbf{S}%
_{n}\cdot \mathbf{S}_{n+2}-h\sum S_{n}^{z}  \label{H}
\end{equation}
where $\mathbf{S}_{n}$ is the spin operator on $n$-th site, $h$ is the
external magnetic field and the exchange integrals are $J_{1}<0$ and $%
J_{2}>0 $.

This model is characterized by the frustration parameter $\alpha
=J_{2}/\left\vert J_{1}\right\vert $. The ground state phase diagram of the
quantum $s=\frac{1}{2}$ model has been intensively studied \cite%
{Chubukov,Itoi,Dmitriev,Vekua,Lu,Richter,Kuzian,Kecke,Sudan}. The ground
state of model (\ref{H}) at $h=0$ is ferromagnetic for $\alpha <\frac{1}{4}$%
. At $\alpha =\frac{1}{4}$ the quantum phase transition to the phase with
incommensurate spin correlations of the helical type takes place.
Remarkably, this transition occurs at the same frustration parameter $\alpha
=\frac{1}{4}$ both in the quantum and in the classical model. However, the
influence of the frustration on the low-temperature thermodynamics in the
vicinity of the transition point is less studied. This problem is of a
special interest because recently studied edge-sharing compound $%
Li_{2}ZrCuO_{4}$ is well described by the F-AF model with the frustration
parameter close to $\alpha =\frac{1}{4}$ \cite{Volkova}.

At present the low-temperature thermodynamics of the quantum $s=\frac{1}{2}$
model at $\alpha \neq 0$ can be studied only either by numerical calculation
of finite chains or by approximate methods. On the other hand, the classical
version of model (\ref{H}) can be studied exactly at $T\to 0$ and the
classical limit is a starting point for the study of quantum effects.
Another reason to study the classical version of F-AF model comes from the
following argument established for the quantum spin-$s$ ferromagnetic chain,
i.e. for model (\ref{H}) at $\alpha =0$. It was conjectured in Ref.\cite%
{Sachdev} that the low-temperature magnetization of this model is a function
of the scaling variable $g_{F}=s^{3}\left\vert J_{1}\right\vert h/T^{2}$.
According to this scaling hypothesis the normalized magnetization $%
m=\left\langle S^{z}\right\rangle /s$ ($\left\langle S^{z}\right\rangle $ is
the magnetization per site) of the quantum chain is expressed as
\begin{equation}
m(T,h)=\phi (g_{F})
\end{equation}

This equation is valid in the scaling limit, which means that $T\to 0$ and $%
h\to 0$ but $g_{F}$ is fixed. Then the dependence of the magnetization $m$
on the spin magnitude $s$ comes only via the scaling variable $g_{F}$.
Generally, the calculation of the function $\phi (g_{F})$ is a very
complicated problem. It was proposed in Ref.\cite{Sachdev} that this
function can be obtained from the solution of the classical ferromagnetic
chain and such scaling function $\phi (g_{F})$ was obtained explicitly in
Ref.\cite{Sachdev,Bacalis}. In particular, the zero-field susceptibility $%
\chi $ is%
\begin{equation}
\chi =s\left. \frac{dm}{dh}\right\vert _{h=0}=\frac{2s^{4}|J_{1}|}{3T^{2}}
\label{chiHF}
\end{equation}

Actually the conjecture of the universality of the function $\phi (g_{F})$
is based on the following observations \cite{Nakamura,Sachdev}: the
zero-field susceptibility of the $s=\frac{1}{2}$ Heisenberg ferromagnetic
chain at $T\to 0$ coincides with that given by Eq.(\ref{chiHF}); the
magnetization $m(T,h)$ obtained numerically from the thermodynamic
Bethe-ansatz equations and plotted as a function of $g_{F}=h/8T^{2}$
approaches to $\phi (g_{F})$ at $T\to 0$; the leading terms of the spin-wave
expansion for magnetization coincide with those for $\phi (g_{F})$. In
addition, as noted in Ref.\cite{Sachdev}, the hypothesis of the universality
originates in the universal behavior of the spin-wave excitations from the
ferromagnetic ground state for both quantum and classical model. For this
reason it is naturally to expect that such universality remains for all $%
\alpha $ corresponding to the ferromagnetic ground state, i.e. for $0\leq
\alpha \leq \frac{1}{4}$. Moreover, the function $\phi (g_{F})$ for $0\leq
\alpha <\frac{1}{4}$ (but $\alpha $ not too close to $\frac{1}{4}$) will be
the same as for $\alpha =0$ but with $g_{F}$ replaced by $g_{F}(1-4\alpha )$%
. Really, the zero-field susceptibility $\chi =2s^{4}(1-4\alpha )/3T^{2}$
fits very well with numerical and analytical results \cite{Hartel}. However,
$\chi T^{2}$ vanishes at $\alpha =\frac{1}{4}$ signalling the change of the
critical exponent at the transition point.

In our previous paper \cite{DK11} we studied the zero-field susceptibility
of the classical F-AF chain exactly at the transition point $\alpha =\frac{1%
}{4}$ and we have shown that $\chi \sim T^{-4/3}$ in contrast with the
low-temperature asymptotic $\chi\sim T^{-2}$ for $0\leq \alpha <\frac{1}{4}$%
. As was shown in Ref.\cite{DK11} the change of the critical exponent for $%
\chi$ is a consequence of the modification of the energy of spin-wave
excitations from $\varepsilon (k)\sim k^{2}$ for $0\leq \alpha <\frac{1}{4}$
to $\varepsilon (k)\sim k^{4}$ at $\alpha =\frac{1}{4}$. Therefore, the form
of the universal magnetization curve and the scaling variable for $\alpha=%
\frac{1}{4}$ (if the universality is valid) differ from the case $0\leq
\alpha <\frac{1}{4}$ and require a special study.

Another interesting problem related to the F-AF model is the influence of
the anisotropy of exchange interactions of the easy-axis type on the
low-temperature magnetic properties of this model. This problem is actual
because it is known that in the real edge-sharing compounds the exchange
interactions are anisotropic and the anisotropy can be of the easy-axis type
\cite{vasiliev, krug}. Though this anisotropy is weak it can be important
especially near the transition point. In particular, it essentially changes
the behavior of the zero-field susceptibility \cite{DK09}.

In this paper we investigate the effect of weak anisotropy on the magnetic
curves of the classical F-AF model at the transition point. In the
low-temperature limit the easy-axis anisotropy of the NN and NNN
interactions have the same effect (we will explain this fact below) and for
simplicity we consider the anisotropy of the NN interaction only, i.e. we
add to Hamiltonian (\ref{H}) the term
\begin{equation}
-(\Delta -1)\sum S_{n}^{z}S_{n+1}^{z}  \label{term}
\end{equation}%
where $\Delta >1$.

It is interesting to note that for the pure ferromagnetic case ($\alpha =0$)
the similarity in the magnetic properties of quantum and classical models
remains in the case of the easy-axis anisotropy. This resemblance is based
on the close relation between the classical solitons and the quantum
multimagnon bound complexes. In this paper we will elucidate the question to
which extend the resemblance between the anisotropic quantum and classical
models remains in the F-AF model.

The paper is organized as follows. In Section II the continuum version of
the model is introduced and the scaling parameters are determined. The
calculation of the partition function is reduced to the solution of the Schr%
\"{o}dinger equation of a special type. In Section III the behavior of the
magnetization curve at the transition point is studied. The asymptotics of
magnetization for low and high magnetic field are presented and relation to
the quantum spin model is discussed. The numerical and analytical results
for the magnetization curve in the helical phase are given in Section IV. In
Section V the influence of the easy-axis anisotropy on the magnetic
properties is studied. The summary of the obtained results is given in
Section VI.

\section{Partition function in the continuum limit}

In Refs.\cite{DK11, DKEPJ} we studied the partition function and the spin
correlation functions of the classical F-AF chain in the vicinity of the
transition point $\alpha =\frac{1}{4}$ at zero magnetic field. This study
was based on the use of a continuum approximation and the interpretation of
the partition function as a path integral for the quantum particle in a
potential well. However, the extension of the model to non-zero magnetic
field and/or non-zero anisotropy needs the essential modification of this
approach.

In the vicinity of the transition point $\alpha =\frac{1}{4}$ it is
convenient to rewrite Hamiltonian (\ref{H}) with the anisotropic term (\ref%
{term}) in the form
\begin{equation}
H=\frac{1}{8}\sum (\mathbf{S}_{n+1}-2\mathbf{S}_{n}+\mathbf{S}_{n-1})^{2}-
\frac{1}{2}(\alpha -\frac{1}{4})\sum (\mathbf{S}_{n+2}-\mathbf{S}
_{n})^{2}-(\Delta -1)\sum S_{n}^{z}S_{n+1}^{z}-h\sum S_{n}^{z}  \label{H2}
\end{equation}

In Eq.(\ref{H2}) we put $\left\vert J_{1}\right\vert =1$ and omit
unessential constant.

In the classical approximation the spin operators $\mathbf{S}_{i}$ are
replaced by the classical vectors $\vec{S}_{i}=s\vec{n}_{i}$, where $\vec{n}%
_{i}$ are the unit vectors. In the low-temperature limit the thermal
fluctuations are weak so that neighbor spins are directed almost parallel to
each other. Therefore, at $T\to 0$ we can use the continuum approximation
replacing $\vec{n}_{i}$ by the classical unit vector field $\vec{n}(x)$, so
that
\begin{eqnarray}
(\vec{S}_{i+1}-2\vec{S}_{i}+\vec{S}_{i-1}) &\simeq &s\frac{\partial ^{2}\vec{%
n}(x)}{\partial x^{2}}  \nonumber \\
(\vec{S}_{i+2}-\vec{S}_{i}) &\simeq &2s\frac{\partial \vec{n}(x)}{\partial x}
\label{continuum}
\end{eqnarray}%
where the lattice constant is chosen as unit length.

Using Eqs.(\ref{continuum}) Hamiltonian (\ref{H2}) goes over into the energy
functional
\begin{equation}
E\left[ \vec{n}(x)\right] =\int dx\left[ \frac{s^{2}}{8}\left( \frac{d^{2}%
\vec{n}}{dx^{2}}\right) ^{2}-\frac{s^{2}(4\alpha -1)}{2}\left( \frac{d\vec{n}%
}{dx}\right) ^{2}-s^{2}(\Delta -1)n_{z}^{2}-shn_{z}\right]  \label{E}
\end{equation}

One can easily check that in the continuum approximation the easy-axis
anisotropy of NNN interactions $\Delta _{2}$ results in the term $\alpha
s^{2}(\Delta _{2}-1)n_{z}^{2}$. This term merely changes the coefficient in
the third term of Eq.(\ref{E}), so that the results obtained below cover the
case of the NNN anisotropy as well.

Energy functional (\ref{E}) contains the second order derivative $d^{2}\vec{n%
}/dx^{2}$ in contrast with that for the ferromagnetic chain \cite{Sachdev},
which contains the first order derivative $d\vec{n}/dx$ only. This fact
demonstrates an essential difference between the cases $\alpha =0$ and $%
\alpha =\frac{1}{4}$.

The partition function is a functional integral over all configurations of
the vector field on a ring of length $L$
\begin{equation}
Z=\int D\left[ \vec{n}(x)\right] \exp \left( -\frac{E\left[ \vec{n}(x)\right]
}{T}\right)  \label{Z}
\end{equation}

It is useful to scale the spatial variable as
\begin{equation}
\xi =\frac{T^{1/3}x}{s^{2/3}}  \label{xi}
\end{equation}

Then, the partition function takes the dimensionless form
\begin{equation}
Z=\int D\left[ \vec{n}(\xi )\right] \exp \left\{ -\int_{0}^{\lambda }d\xi %
\left[ \frac{1}{8}\left( \frac{d^{2}\vec{n}}{d\xi ^{2}}\right) ^{2}-\frac{%
\gamma }{2}\left( \frac{d\vec{n}}{d\xi }\right) ^{2}-\delta n_{z}^{2}-gn_{z}%
\right] \right\}  \label{Z2}
\end{equation}
where
\begin{equation}
\lambda =\frac{T^{1/3}L}{s^{2/3}}  \label{lambda}
\end{equation}
is the scaled system length and
\begin{equation}
\gamma =\frac{(4\alpha -1)s^{4/3}}{T^{2/3}},\quad \delta =\frac{(\Delta
-1)s^{8/3}}{T^{4/3}},\quad g=\frac{hs^{5/3}}{T^{4/3}}  \label{scpar}
\end{equation}
are the parameters of the model scaled by temperature.

As follows from Eq.(\ref{Z2}) the partition function and with it the
low-temperature thermodynamics of the F-AF model near the transition point
is governed by three scaling parameters $\gamma $, $\delta $ and $g$. The
definition of the scaling parameters (\ref{scpar}) implies that we consider
the scaling limit when $T\to 0$, $\alpha \to \frac{1}{4}$, $\Delta \to 1$, $%
h\to 0$, but the values of the scaling parameters $\gamma$, $\delta$ and $g$
are finite.

We express the unit vector field through two scalar fields $\phi (\xi )$ and
$n_{z}(\xi )$
\begin{equation}
\vec{n}(\xi )=(\cos \phi \sqrt{1-n_{z}^{2}},\sin \phi \sqrt{1-n_{z}^{2}}%
,n_{z})
\end{equation}%
and the magnetic field is directed along the $Z$ axis. In terms of the
fields $n_{z}(\xi )$ and $\phi (\xi )$ the partition function takes the form
of the functional integral
\begin{equation}
Z=\int D\left[ n_{z}(\xi )\right] D\left[ \phi (\xi )\right] \exp \left\{
-\int_{0}^{\lambda }W(n_{z},\phi )d\xi \right\}  \label{Z3d}
\end{equation}%
where the energy density $W(n_{z},\phi )$ has a rather cumbersome form:
\begin{eqnarray}
W &=&\frac{1}{8}\left[ (1-n_{z}^{2})\left( \ddot{\phi}^{2}+\dot{\phi}%
^{4}\right) +\frac{\ddot{n}_{z}^{2}}{1-n_{z}^{2}}+\frac{\dot{n}_{z}^{4}}{%
(1-n_{z}^{2})^{3}}-4n_{z}\dot{n}_{z}\dot{\phi}\ddot{\phi}+\frac{2n_{z}\dot{n}%
_{z}^{2}\ddot{n}_{z}}{(1-n_{z}^{2})^{2}}+2n_{z}\ddot{n}_{z}\dot{\phi}^{2}+%
\frac{2+4n_{z}^{2}}{1-n_{z}^{2}}\dot{n}_{z}^{2}\dot{\phi}^{2}\right]
\nonumber \\
&&-\frac{\gamma }{2}\left[ \frac{\dot{n}_{z}^{2}}{1-n_{z}^{2}}+(1-n_{z}^{2})%
\dot{\phi}^{2}\right] -\delta n_{z}^{2}-gn_{z}  \label{W}
\end{eqnarray}

Here $\dot{n}_{z}$, $\ddot{n}_{z}$ and $\dot{\phi}$, $\ddot{\phi}$ are the
first and the second-order derivatives of $n_{z}$ and $\phi $ with respect
to $\xi $.

If we treat $\xi $ as an imaginary time then partition function (\ref{Z3d})
takes the form of a path integral of a quantum particle with the Euclidean
Lagrangian $W(n_{z},\phi )$. Here we notice that $W(n_{z},\phi )$ comprises
the derivatives $\dot{\phi}$ and $\ddot{\phi}$ of the field $\phi (\xi )$,
but does not contain explicitly the field $\phi (\xi )$ itself. This allows
us to rewrite partition function (\ref{Z3d}) in terms of a new field
\begin{equation}
q(\xi )=\frac{d\phi }{d\xi },\;\dot{q}(\xi )=\frac{d^{2}\phi }{d\xi ^{2}}
\end{equation}

The energy density $W(n_{z},\phi )$ contains explicitly the field $n_{z}(\xi
)$ and its derivatives $\dot{n}_{z}$ and $\ddot{n}_{z}$. Presence of the
second-order derivative requires the use of the special methodology
developed by Ostrogradski \cite{ostrog} which allows to obtain the
Hamiltonian corresponding to the higher gradient Lagrangian. In the
Ostrogradski formalism \cite{Kleinert} the independent generalized
coordinates are $n_{z}$ and $v=\dot{n}_{z}$. That is, we treat the
derivative $\dot{n}_{z}=v$ as a new independent variable, so that
\begin{equation}
v(\xi )=\frac{dn_{z}}{d\xi },\;\dot{v}(\xi )=\frac{d^{2}n_{z}}{d\xi ^{2}}
\end{equation}

According to this formalism the Lagrangian (\ref{W}) is replaced by the
equivalent one
\begin{eqnarray}
L &=&\frac{1}{8}\left[ (1-n_{z}^{2})(\dot{q}^{2}+q^{4})+\frac{\dot{v}^{2}}{%
(1-n_{z}^{2})}+\frac{v^{4}}{(1-n_{z}^{2})^{3}}-4n_{z}vq\dot{q}+\frac{%
2n_{z}v^{2}\dot{v}}{(1-n_{z}^{2})^{2}}+2n_{z}\dot{v}q^{2}+\frac{2+4n_{z}^{2}%
}{1-n_{z}^{2}}q^{2}v^{2}\right]  \nonumber \\
&&-\frac{\gamma }{2}\left[ \frac{v^{2}}{1-n_{z}^{2}}+(1-n_{z}^{2})q^{2}%
\right] -\delta n_{z}^{2}-gn_{z}-ip(\dot{n}_{z}-v)  \label{L3}
\end{eqnarray}%
where the Lagrange multiplier $p$ ensures the equality of $\dot{n}_{z}$ and $%
v$. The canonical momenta are $p=i\frac{\partial \tilde{L}}{\partial \dot{n}%
_{z}}$, $p_{v}=i\frac{\partial \tilde{L}}{\partial \dot{v}}$ and $p_{q}=i%
\frac{\partial \tilde{L}}{\partial \dot{q}}$.

Then, partition function (\ref{Z3d}) takes the form written in terms of
three scalar fields $q(\xi )$, $v(\xi )$ and $n_z(\xi )$:
\begin{equation}
Z=\int D\left[ n_{z}\right] D\left[ v\right] D\left[ q\right] \exp \left\{
-\int_{0}^{\lambda }L(n_{z},v,q)d\xi \right\}  \label{Z4}
\end{equation}

The last term in Eq.(\ref{L3}) is a specific property of the Ostrogrdski
methodology and we will pay a special attention to it because it makes the
following quantum Hamiltonian a non-Hermitian one.

Now we construct the Hamilton function $H=ip\dot{n}_{z}+ip_{v}\dot{v}+ip_{q}%
\dot{q}+L$, which after replacing momenta by the corresponding differential
operators: $\hat{p}=-i\frac{\partial }{\partial n_{z}}$, $\hat{p}_{v}=-i%
\frac{\partial }{\partial v}$ and $\hat{p}_{q}=-i\frac{\partial }{\partial q}
$ results in the quantum Hamiltonian:
\begin{eqnarray}
\hat{H} &=&-2(1-n_{z}^{2})\frac{\partial ^{2}}{\partial v^{2}}-\frac{2}{%
1-n_{z}^{2}}\frac{\partial ^{2}}{\partial q^{2}}-n_{z}\left( \frac{v^{2}}{%
1-n_{z}^{2}}+(1-n_{z}^{2})q^{2}\right) \frac{\partial }{\partial v}+\frac{%
2n_{z}vq}{1-n_{z}^{2}}\frac{\partial }{\partial q}+v\frac{\partial }{%
\partial n_{z}}  \nonumber \\
&&+\frac{1}{8}\left( \frac{v^{2}}{1-n_{z}^{2}}+(1-n_{z}^{2})q^{2}\right)
^{2}-\frac{\gamma }{2}\left( \frac{v^{2}}{1-n_{z}^{2}}+(1-n_{z}^{2})q^{2}%
\right) -\delta n_{z}^{2}-gn_{z}  \label{H3d}
\end{eqnarray}

It is convenient to change variables $v$, $q$, $n_{z}$ to new variables $r$,
$\varphi $, $\theta $ connected by the relations
\begin{equation}
v=r\cos \varphi \sin \theta ,\;q=\frac{r\sin \varphi }{\sin \theta }%
,\;n_{z}=\cos \theta
\end{equation}

Then we obtain the Schr\"{o}dinger equation for the quantum particle in the
form
\begin{equation}
\hat{H}_{0}\Psi _{n}-\frac{\gamma }{2}r^{2}\Psi _{n}-g\cos \theta \Psi
_{n}-\delta \cos ^{2}\theta \Psi _{n}=\varepsilon _{n}\Psi _{n}
\label{sch3d}
\end{equation}
where
\begin{equation}
\hat{H}_{0}=-2\left( \frac{\partial ^{2}}{\partial r^{2}}+\frac{1}{r}\frac{%
\partial }{\partial r}+\frac{1}{r^{2}}\frac{\partial ^{2}}{\partial \varphi
^{2}}\right) +\frac{1}{8}r^{4} +r\sin \varphi \cot \theta\frac{\partial}{%
\partial \varphi } -r\cos \varphi \frac{\partial }{\partial \theta }
\label{H0}
\end{equation}
describes the model at the transition point at $h=0$.

The last two terms in Eq.(\ref{H0}) makes the Hamiltonian to be
non-Hermitian one. Therefore, we have to consider the transposed counterpart
of Eq.(\ref{sch3d}):
\begin{equation}
\hat{H}_{0}^{T}\Phi _{n}-\frac{\gamma }{2}r^{2}\Phi _{n}-g\cos \theta \Phi
_{n}-\delta \cos ^{2}\theta \Phi _{n}=\varepsilon _{n}\Phi _{n}  \label{con}
\end{equation}

The transposed differential operator $\hat{H}_{0}^{T}$ has the same form as $%
\hat{H}_{0}$, but the sign of the last two terms in Eq.(\ref{H0}) is
changed. This change of sign is equivalent to the change $\theta \to -\theta
$, which implies that $\Phi _{n}(r,\theta ,\varphi )=\Psi _{n}(r,-\theta
,\varphi )$. Then, the normalization condition for functions $\Psi _{n}$
takes the form:
\begin{equation}
\frac{1}{4\pi }\int_{0}^{\infty }rdr\int_{0}^{\pi }\sin \theta d\theta
\int_{0}^{2\pi }d\varphi \Psi _{n}(r,\theta ,\varphi )\Psi _{m}(r,-\theta
,\varphi )=\delta _{nm}  \label{n}
\end{equation}

As a result of the above manipulations the partition function $Z$ can be
considered as the partition function of quantum model (\ref{sch3d}) at a
`temperature' $1/\lambda $%
\begin{equation}
Z=\sum e^{-\lambda \varepsilon _{n}}
\end{equation}

In the thermodynamic limit $\lambda \to \infty $ ($\lambda =s^{-2/3}T^{1/3}L$%
) only the lowest eigenvalue of Eq.(\ref{sch3d}) gives the contribution to $Z
$. Thus, the free energy of the classical spin model is determined by the
ground state energy $\varepsilon _{0}$ of the Schr\"{o}dinger equation (\ref%
{sch3d}). The dependence of the lowest eigenvalue $\varepsilon _{0}$ of Eq.(%
\ref{sch3d}) on the scaling parameters $g$, $\gamma $ and $\delta $
determines the magnetic properties of the system. In particular, the
normalized magnetization is given by
\begin{equation}
m=-\frac{\partial \varepsilon _{0}}{\partial g}  \label{magn}
\end{equation}

Thus, Eq.(\ref{sch3d}) is the main result of this paper. In general, Eq.(\ref%
{sch3d}) does not admit analytical solution and should be solved
numerically. However, the limiting cases of high and low magnetic fields can
be studied analytically. In the following we present both numerical
solutions and analytical expressions for asymptotics.

\section{Magnetization curve at the transition point}

At first, let us consider the isotropic F-AF model at the transition point
when $\delta =0$ and $\gamma =0$. For low magnetic field ($g\ll 1$) the
ground state energy $\varepsilon _{0}$ can be found using the PT in $g$. The
numerical solution of Eq.(\ref{sch3d}) for $g=0$ shows that the ground state
wave function does not depend on $\varphi $ and $\theta $, i.e. it satisfies
the equation
\begin{equation}
-2\frac{\partial ^{2}\Psi _{0}}{\partial r^{2}}-\frac{2}{r}\frac{\partial
\Psi _{0}}{\partial r}+\frac{1}{8}r^{4}\Psi _{0}=\varepsilon _{0}\Psi _{0}
\end{equation}
and $\varepsilon _{0}=1.861$.

The eigenfunctions of Eq.(\ref{sch3d}) giving the contribution to the second
order in $g$ have the form
\begin{equation}
\Psi _{n}=f_{1n}(r)\cos \varphi \sin \theta +f_{2n}(r)\cos \theta
\end{equation}
where the functions $f_{1n}(r)$ and $f_{2n}(r)$ satisfy the following system
of equations:
\begin{eqnarray}
-2\frac{\partial ^{2}f_{1n}}{\partial r^{2}}-\frac{2}{r}\frac{\partial f_{1n}%
}{\partial r}+\frac{2}{r^{2}}f_{1n}+\frac{1}{8}r^{4}f_{1n}-rf_{2n}
&=&\varepsilon _{n}f_{1n}  \nonumber \\
-2\frac{\partial ^{2}f_{2n}}{\partial r^{2}}-\frac{2}{r}\frac{\partial f_{2n}%
}{\partial r}+\frac{1}{8}r^{4}f_{2n}+rf_{1n} &=&\varepsilon _{n}f_{2n}
\label{eeqq}
\end{eqnarray}

Normalization condition (\ref{n}) transforms for the functions $f_{1n}(r)$
and $f_{2n}(r)$ to equation
\begin{equation}
\frac{1}{3}\int_{0}^{\infty }rdr(f_{2n}f_{2m}-f_{1n}f_{1m})=\delta _{nm}
\end{equation}

Further, we calculate the second-order correction to the ground state energy
in $g$:
\begin{equation}
\varepsilon =\varepsilon _{0}+\frac{g^{2}}{9}\sum_{n}\frac{M_{0n}^{2}}{%
\varepsilon _{0}-\varepsilon _{n}}  \label{ee}
\end{equation}
where $M_{0n}$ is the following matrix element
\begin{equation}
M_{0n}=\int_{0}^{\infty }\Psi _{0}(r)f_{2n}(r)rdr
\end{equation}

The numerical solution of Eq.(\ref{eeqq}) and the calculation of the sum in
Eq.(\ref{ee}) gives
\begin{equation}
\varepsilon =\varepsilon _{0}-0.534g^{2}
\end{equation}

Then, the magnetization $m$ at $g\to 0$ is
\begin{equation}
m=1.07g+O(g^{3})
\end{equation}
and the zero-field susceptibility is
\begin{equation}
\chi =\frac{1.07s^{8/3}}{T^{4/3}}  \label{chi3d}
\end{equation}

Expression (\ref{chi3d}) naturally reproduces the result found in Ref.\cite%
{DK11} obtained by another method and confirmed by Monte-Carlo simulations
\cite{Sirker}. As follows from Eq.(\ref{chi3d}) the critical exponent of $%
\chi $ is changed from $2$ to $\frac{4}{3}$ when $\alpha \to \frac{1%
}{4}$ from the ferromagnetic side.

If we assume that the hypothesis of the universality is valid, then the
susceptibility $\chi $ at the transition point for the $s=\frac{1}{2}$ F-AF
chain at $T\to 0$ is
\begin{equation}
\chi =\frac{0.1681}{T^{4/3}}  \label{chi12}
\end{equation}

Unfortunately, the exact low-temperature asymptotic of $\chi $ for the $s=%
\frac{1}{2}$ F-AF model at $\alpha =\frac{1}{4}$ is unknown. However, we can
compare Eq.(\ref{chi12}) with the susceptibility obtained for this model by
the approximate modified spin-wave method (MSWT) proposed by Takahashi \cite%
{Takahashi}. The MSWT gives $\chi =0.099T^{-4/3}$
\cite{DK09,Sirker}. The comparison of MSWT result with
Eq.(\ref{chi12}) shows that the critical exponents of both
expressions are the same though the prefactors are different. In
Ref.\cite{Sirker} the transfer-matrix renormalization group (TMRG)
algorithm was used for the calculation of the low-temperature
asymptotic of $\chi $. The obtained numerical results are not fully
consistent with Eq.(\ref{chi12}) and show that the exponent might
actually be smaller than $4/3$. However, as pointed in
Ref.\cite{Sirker} the possible reason of the deviation of the TMRG
results from Eq.(\ref{chi12}) is that the obtainable temperatures in
the TMRG calculations are just not low enough to observe the
$T^{-4/3}$ power law predicted by Eq.(\ref{chi12}).

Now we consider the limit of large $g$ when the magnetization is close to
saturation. In this limit we expand $\cos \theta $ near $\theta =0$ and
scale the variables $r$ and $\theta $ as:
\begin{equation}
r=zg^{-1/8},\;\theta =xg^{-3/8}  \label{scal}
\end{equation}

Keeping in Eq.(\ref{sch3d}) the terms proportional to $g^{1/4}$ we arrive at
the Schr\"{o}dinger equation in a form
\begin{equation}
-2\left( \frac{\partial ^{2}\Psi }{\partial z^{2}}+\frac{1}{z}\frac{\partial
\Psi }{\partial z}+\frac{1}{z^{2}}\frac{\partial ^{2}\Psi }{\partial \varphi
^{2}}\right) -\frac{z\sin \varphi }{x}\frac{\partial \Psi }{\partial \varphi
}+z\cos \varphi \frac{\partial \Psi }{\partial x}+\frac{x^{2}}{2}\Psi =\frac{%
\varepsilon +g}{g^{1/4}}\Psi  \label{gi}
\end{equation}

Fortunately, the ground state wave function and the ground state energy of
Eq.(\ref{gi}) can be found exactly:
\begin{eqnarray}
\Psi _{0}(x,z,\varphi ) &=&C\exp \left( -\frac{z^{2}}{4}+\frac{zx\cos
\varphi }{2}-\frac{x^{2}}{2}\right)  \label{psi0} \\
\varepsilon _{0} &=&-g+2g^{1/4}  \label{eg}
\end{eqnarray}
where $C$ is unessential normalization constant.

One can also calculate the next-order correction to the ground state energy (%
\ref{eg}). For this aim we estimate the effect of the next-order term which
was omitted in Eq.(\ref{gi}) and has the form:
\begin{equation}
g^{-3/4} \left(\frac{3z^{4}-x^{4}}{24}+\frac{xz }{3}\sin \varphi\frac{%
\partial}{\partial \varphi }\right)  \label{per}
\end{equation}

The calculation of the first order in perturbation (\ref{per}) gives for $%
\varepsilon _{0}(g)$ the correction proportional to $g^{-1/2}$:
\begin{equation}
\varepsilon _{0}=-g+2g^{1/4}+\frac{3}{4g^{1/2}}  \label{ept}
\end{equation}

Then, the asymptotics for the magnetization and the susceptibility for $h\gg
T^{4/3}$ are
\begin{eqnarray}
m &=&1-\frac{1}{2g^{3/4}}+\frac{3}{8g^{3/2}}+O(g^{-9/4})  \label{mtp} \\
\chi (h) &=&\frac{3T}{8h^{7/4}s^{1/4}}\left[ 1-\frac{3}{2}\frac{T}{%
h^{3/4}s^{5/4}}+O\left( \frac{T^{2}}{h^{3/2}s^{5/2}}\right) \right]
\end{eqnarray}

It is interesting to compare the leading terms of this expansion with the
spin-wave expansion of the magnetization for the spin-$s$ quantum F-AF chain
at $\alpha =\frac{1}{4}$. We have checked that this expansion reproduces Eq.(%
\ref{mtp}). The second term in Eq.(\ref{mtp}) corresponds to the result of
the linear spin-wave theory, but the third one includes spin-wave
interaction effect and, therefore, the coincidence is not trivial.
Certainly, we can not prove that both expansions coincide in all orders in
small parameter $g^{-3/4}$. Nevertheless, the coincidence of the leading
terms of $m(g)$ for the quantum and the classical model gives a promise that
the universality is valid at the transition point of the F-AF model.

We complete this subsection with the results for the spin correlation
function. It can be shown \cite{DK11} that the spin correlation function $%
\left\langle S_{z}(0)S_{z}(l)\right\rangle $ has the form
\begin{equation}
\left\langle S_{z}(0)S_{z}(l)\right\rangle =\sum \left\langle \Psi _{0}|\Psi
_{n}\right\rangle ^{2}\exp [-T^{1/3}(\varepsilon _{n}-\varepsilon _{0})l]
\label{cor}
\end{equation}

As follows from Eq.(\ref{cor}) the spin correlation function exponentially
decays on long distances $l\gg T^{-1/3}$, and the correlation length is
governed by the lowest eigenstates of Eq.(\ref{sch3d}). In the case of
absence of the magnetic field ($g=0$) all the eigenvalues $\varepsilon _{n}$
are real and several lowest levels was calculated in \cite{DK11}, which
gives the asymptotic for the correlation length $l_{0}=1.04T^{-1/3}$ at $%
g\to 0$.

In the high field limit ($g\gg 1$) there are three lowest excited states
having equal real part of their eigenvalues:
\begin{eqnarray}
\varepsilon _{1} &=&-g+4g^{1/4}+O(g^{-1/2})  \nonumber \\
\varepsilon _{2,3} &=&-g+2g^{1/4}(2\pm i)+O(g^{-1/2})  \label{e123}
\end{eqnarray}

According to Eq.(\ref{cor}) the presence of the imaginary part in
eigenvalues (\ref{e123}) causes the oscillations on the background of the
exponential decay of the correlation function. The imaginary part of the
eigenvalues determines the period of the oscillations while the real part
determines the correlation length. According to Eqs.(\ref{ept}) and (\ref%
{e123}) the asymptotic of the correlation length in the high field limit is
\begin{equation}
l_{0}=\frac{1}{2s^{5/12}h^{1/4}}  \label{l0}
\end{equation}

So we see that the correlation length is defined by the temperature for $%
h\ll T^{4/3}$ and by the magnetic field when $h\gg T^{4/3}$. We note that
the ratio of the correlation lengths in these limits is proportional to $%
g^{1/4}$. The crossover between these two regimes occurs at $g\approx 1$.

\begin{figure}[tbp]
\includegraphics[width=3in,angle=0]{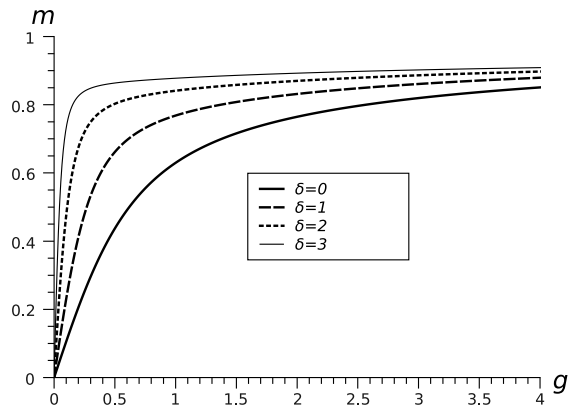}
\caption{Magnetization curves $m(g)$ at $\gamma=0$ for some values
of scaled anisotropy $\delta$. The case $\delta=0$ corresponds to
the transition point.} \label{delta}
\end{figure}

In general, the solution of Eq.(\ref{sch3d}) and the computation of $%
\varepsilon _{0}(g)$ and $m(g)$ has been obtained numerically. The
dependence $m(g)$ at the transition point is shown by thick solid line in
Fig.\ref{delta}.

\section{Magnetization curve in the helical phase}

In this section we consider the behavior of the magnetization in the helical
phase in the vicinity of the transition point, when $\alpha >\frac{1}{4}$,
but the anisotropy is zero $\delta=0$. For $\alpha >\frac{1}{4}$ the ground
state has the helical type of long-range order (LRO) with the wave-number $%
k_{h}=\cos^{-1}(1/4\alpha )$. The saturation field $h_{s}$ at $\alpha $
close to $\frac{1}{4}$ is $h_{s}=s(4\alpha -1)^{2}$. The ground state
magnetization is given by $m=h/h_{s}$ for $h\leq h_{s}$ and $m=1$ for $%
h>h_{s}$. At finite temperature the helical LRO is destroyed by thermal
fluctuations and thermodynamic quantities have singular behavior at $T\to 0$.

The behavior of the system in the case of absence of the magnetic field was
studied in detail in Ref.\cite{DKEPJ}. It was shown that with the increase
of the temperature the gapless excitations over the helical ground states
(spin waves) smear the $\delta $-peaks of the static structure factor at $%
k=\pm k_{h}$ and shift the peaks to $k=0$. Finally, at $T_{c}=0.925\gamma
^{3/2}$ the maximum of the spin structure factor reaches $k=0$ defining the
Lifschitz boundary, so that the helical type of the spin correlations for $%
T>T_{c}$ disappears.

Most likely, the hypothesis of the universality of the function $m(g)$
breaks down for $\alpha >\frac{1}{4}$ because the excitations above the
ground state are different in the quantum and in the classical F-AF chain.
Nevertheless, as was shown in Ref.\cite{DK11} some peculiarities of the
low-temperature behavior of the classical model at $\alpha >\frac{1}{4}$ is
qualitatively similar to that for the quantum $s=\frac{1}{2}$ chain. For
example, the temperature dependence of the zero-field susceptibility is in a
qualitative agreement with the numerical data for the quantum $s=\frac{1}{2}$
model and is in accord with the experimental data for the real edge-sharing
compounds.

The finding of the magnetization in the helical phase reduces to the
solution of Eq.(\ref{sch3d}) for $\gamma >0$ and can be analyzed in full
analogy with the case $\gamma =0$. For low magnetic field ($g\to 0$) the
magnetization is $m\sim g$ and can be represented as
\begin{equation}
m=\frac{h}{h_{s}}G(\gamma )
\end{equation}

The function $G(\gamma )$ is found from the solution of Eq.(\ref{eeqq})
where the terms $-\frac{1}{2}\gamma r^{2}f_{1(2)n}$ are added to the first
(the second) equation of (\ref{eeqq}). In fact, $G(\gamma)$ coincides with
the normalized zero-field susceptibility obtained before in Ref.\cite{DK11}.
Therefore, we do not present this function here. We note only that $%
G(\gamma) $ vanishes at $\gamma\to\infty$, tends to finite value at $%
\gamma\to 0$ and has a maximum at $\gamma\simeq 2.2$.

In high magnetic field limit we use rescaling (\ref{scal}) for Eq.(\ref%
{sch3d}) and keep the leading terms. The obtained equation repeats Eq.(\ref%
{gi}) with the additional term $-\frac{1}{2}\sqrt{h_{s}/h}z^{2}\Psi $. The
ground state wave function of this equation has the form similar to Eq.(\ref%
{psi0}):
\begin{equation}
\Psi =C\exp \left( -az^{2}+\frac{1}{2}zx\cos \varphi -2ax^{2}\right)
\label{psia}
\end{equation}
and the ground state energy is
\begin{equation}
\varepsilon_0=-g+8ag^{1/4}
\end{equation}
where
\begin{equation}
a=\frac{1}{4}\sqrt{1-\sqrt{h_{s}/h}}
\end{equation}

Eq.(\ref{psia}) is valid for high fields when $h>h_{s}$. The asymptotic of
the magnetization curve in this limit has the form
\begin{equation}
m=1-\frac{(h_s/h)^{3/4}}{2\gamma^{3/2}\sqrt{1-\sqrt{h_{s}/h}}}  \label{mhs}
\end{equation}

As follows from Eq.(\ref{mhs}) the temperature-dependent correction to $m=1$
at $h\gg h_s$ is proportional to $\gamma^{-3/2}$.

\begin{figure}[tbp]
\includegraphics[width=3in,angle=0]{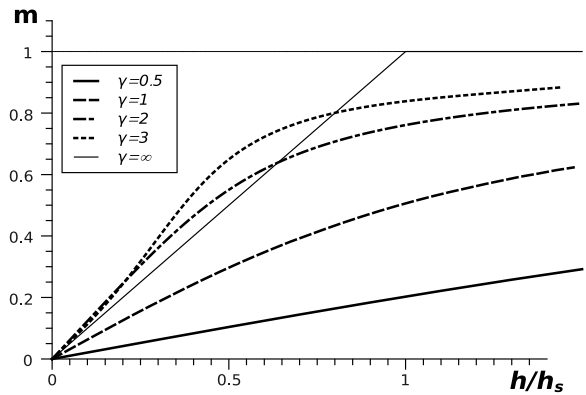}
\caption{Magnetization as a function of $h/h_{s}$ for several values
of parameter $\gamma$ for isotropic case ($\delta=0$).}
\label{m_hhs}
\end{figure}

The magnetization curves for several values of $\gamma$ as a function of $%
h/h_{s}$ are shown in Fig.\ref{m_hhs} together with the ground state
magnetization ($\gamma\to\infty$). The zero-field susceptibility defines the
slopes of the magnetization curves for small $h/h_{s}$ and as follows from
Fig.\ref{m_hhs} this slope can be both larger and smaller than the ground
state value. Such behavior of the magnetization follows from the
non-monotonic dependence of $G$ on $\gamma$.

\section{Easy-axis anisotropy}

Up to now we considered the isotropic F-AF chain. At the same time it is
important to study the influence of the anisotropy of exchange interactions
on the low-temperature thermodynamics. In this subsection we pay our
attention mainly to the dependence of the zero-field susceptibility on the
anisotropy at $\alpha =\frac{1}{4}$.

At first, we briefly review the effect of the anisotropy on the
susceptibility for the classical and quantum ferromagnetic model ($\alpha =0$%
). At $T\to 0$ and in the scaling limit the magnetization of the classical
ferromagnetic chain is given by Eq.(\ref{magn}), where $\varepsilon _{0}$ is
the lowest eigenvalue of the equation \cite{Sachdev,Bacalis}:
\begin{equation}
-\frac{1}{2}\frac{d^{2}\Psi }{d\theta ^{2}}-\frac{1}{2}\cot \theta \frac{%
d\Psi }{d\theta }-\delta _{F}\cos ^{2}\theta \Psi -g_{F}\cos \theta \Psi
=\varepsilon \Psi  \label{HF}
\end{equation}

In this equation $\delta _{F}=(\Delta -1)s^{4}/T^{2}$ and $%
g_{F}=s^{3}h/T^{2} $ are the scaling parameters.

To find the susceptibility in the limit $\delta _{F}\ll 1$ and $g_{F}\ll 1$
we can use the PT in $V=-\delta _{F}\cos ^{2}\theta -g_{F}\cos \theta $. At $%
V=0$ the eigenfunctions of Eq.(\ref{HF}) are the Legendre polynomials $%
P_{l}(\cos \theta )$ with the eigenvalues $\varepsilon _{l}=l(l+1)/2$. The
PT in the lowest orders in $V$ gives
\begin{equation}
\varepsilon _{0}=-\frac{\delta _{F}}{3}-\frac{g_{F}^{2}}{3}-\frac{%
44g_{F}^{2}\delta _{F}}{135}
\end{equation}

Then, according to Eq.(\ref{magn}) the zero-field susceptibility in the
limit of weak anisotropy $\delta _{F}\to 0$ ($\Delta -1\ll T^{2}$) is
\begin{equation}
\chi =\frac{2s^{4}}{3T^{2}}\left( 1+\frac{44s^{4}(\Delta -1)}{45T^{2}}\right)
\label{chiH}
\end{equation}

In the opposed limit $\delta _{F}\to \infty $ and $g_{F}=0$ Eq.(\ref{HF})
has two almost degenerated lowest eigenvalues corresponding to the states
with even and odd parity with respect to exchange $\theta \leftrightarrow
(\pi -\theta )$. The tunnel splitting between these states can be found with
the exponential accuracy in the WKB approximation \cite{Landau}: $\Delta
\varepsilon \sim \exp (-2\sqrt{2\delta _{F}})$. The term ($-g_{F}\cos \theta
$) in Eq.(\ref{HF}) has non-zero matrix element between the states with even
and odd parities, so that the contribution to the second order PT in $g_{F}$
is given by
\begin{equation}
\varepsilon _{0}\sim -g_{F}^{2}\exp \left( 2\sqrt{2\delta _{F}}\right)
\end{equation}
and the zero-field susceptibility is
\begin{equation}
\chi \sim \frac{1}{T^{2}}\exp \left( \frac{\Delta E}{T}\right)  \label{chie}
\end{equation}
where $\Delta E=2s^{2}\sqrt{2(\Delta -1)}$.

As follows from Eq.(\ref{chie}) the susceptibility diverges exponentially at
$T\to 0$ and the value of the thermal gap $\Delta E$ is equal to the kink
energy (or one-half of the energy of large soliton) of the
weakly-anisotropic ferromagnetic chain. As it is known \cite{KIK} the
soliton energy coincides with the energy of the multimagnon bound states of
the quantum anisotropic ferromagnetic chain. It is interesting to note that
the susceptibility of the easy-axis anisotropic $s=\frac{1}{2}$
ferromagnetic chain found on the base of the Gaudin formalism in Ref.\cite%
{Johnson} behaves at $T\to 0$ as $\chi \sim \exp (\Delta E/T)/T$ and $\Delta
E$ is the same as given in Eq.(\ref{chie}). This fact manifests the close
relation between the magnetic properties of the quantum and the classical
anisotropic ferromagnetic chains.

We will show that this resemblance remains in the F-AF model at the critical
point $\alpha =\frac{1}{4}$. For example, it was shown by us in Refs.\cite%
{DK09,DK10} that the energies of multimagnon bound states in the quantum
model and the energy of large classical solitons at $\alpha =\frac{1}{4}$
are both proportional to $(\Delta -1)^{3/4}$ though the numerical
coefficients are different. We have also shown \cite{DK09} that the
susceptibility of the quantum $s=\frac{1}{2}$ F-AF model diverges at $T\to 0$
exponentially, i.e. $\chi \sim \exp (\Delta E/T)$ and the thermal gap $%
\Delta E$ equals one-half of the energy of the multimagnon complexes. We
will show below that the susceptibility of the classical model at $\delta
\gg 1$ has similar exponential behavior and the corresponding thermal gap is
the classical kink energy.

Similar to the pure ferromagnetic chain the calculation of the zero-field
susceptibility reduces to the computation of the ground state energy of Eq.(%
\ref{sch3d}) in the second order in $V=-g\cos \theta $. Then the
susceptibility can be represented as $\chi =2(s^{2}/T)^{4/3}f(\delta )$
where
\begin{equation}
f(\delta )=\sum_{n\neq 0}\frac{\left\langle \Psi _{0}\right\vert \cos \theta
\left\vert \Psi _{n}\right\rangle ^{2}}{\varepsilon _{n}-\varepsilon _{0}}
\label{fd}
\end{equation}
and $\Psi _{n}$ and $\varepsilon _{n}$ are the eigenfunctions and the
eigenvalues of Eq.(\ref{sch3d}) at $g=0$.

It is convenient to introduce the normalized susceptibility $\widetilde{\chi
}=(\Delta -1)\chi $ and the normalized temperature $\tilde{T}=T/s^{2}(\Delta
-1)^{3/4}$ ($\tilde{T}=\delta ^{-3/4}$), so that $\widetilde{\chi }$ is a
function of $\tilde{T}$ only. The function $\widetilde{\chi }(\tilde{T})$
can be found explicitly in the limits of large and small values of $\tilde{T}
$ (at small and large $\delta $ correspondingly). For high temperatures $%
\tilde{T}\gg 1$ the states giving the contributions to the sum in Eq.(\ref%
{fd}) are separated from the ground state by finite gap and the numerical
calculation of this sum gives
\begin{equation}
\widetilde{\chi }=1.07\tilde{T}^{-4/3}+0.87\tilde{T}^{-8/3}  \label{chis}
\end{equation}

The calculation $\widetilde{\chi }(\tilde{T})$ for small $\tilde{T}$ is more
complicated. At $\tilde{T}\to 0$ ($\delta \to \infty $) we can expand the
term $\delta \cos ^{2}\theta $ up to $\theta ^{2}$ in the Schr\"{o}dinger
equation (\ref{sch3d}). Then the ground state wave function has the form
similar to Eq.(\ref{psi0}) and the energy $\varepsilon _{0}=-\tilde{T}%
^{-4/3}+2^{5/4}\tilde{T}^{-1/3}$. Further, we can compute perturbative
corrections to $\varepsilon _{0}$ from omitted anharmonic terms to obtain
the ground state energy in a form
\begin{equation}
E_{0}=\varepsilon _{0}+\sum a_{n}\tilde{T}^{n+2/3}  \label{ep}
\end{equation}

\begin{figure}[tbp]
\includegraphics[width=3in,angle=0]{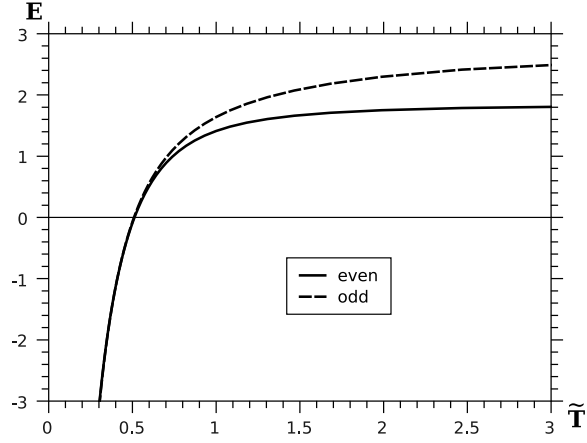}
\caption{Dependence of two lowest eigenvalues of Eq.(\ref{sch3d}) on
the normalized temperature $\tilde{T}=T/s^2(\Delta -1)^{3/4}$ at $%
\gamma=g=0$.}
\label{e_delta}
\end{figure}

If we expand the term $\delta\cos^2\theta$ near the second minimum $\theta
=\pi$ we would obtain the result identical to Eq.(\ref{ep}), so that we have
two degenerated states. But there is a non-perturbative tunnel splitting
which is not captured by the PT. The splitting is exponentially small at $%
\tilde{T}\to 0$ as demonstrated in Fig.\ref{e_delta}. On the other hand,
these quasi-degenerated states have a non-zero matrix element in Eq.(\ref{fd}%
) and, therefore, the splitting between them determines the behavior of $%
\widetilde{\chi }$ at $\tilde{T}\to 0$.

The most convenient way to evaluate this splitting is the calculation of the
original functional integral (\ref{Z3d}). Certainly, the exact calculation
of this integral is impossible and, therefore, we use a semiclassical
approximation. In this approximation the functional integral (\ref{Z3d}) is
represented as the sum of the contributions of the classical paths in $Z$
minimizing the Euclidean action and the paths which are close to the
classical ones. The minimization of the energy functional $W(\theta ,\varphi
)$ (Eq.(\ref{W})) gives $\varphi =0$ and the following Euler equation for $%
\theta (\xi )$:
\begin{equation}
\frac{1}{4}\theta ^{^{\prime \prime \prime \prime }}-\frac{3}{2}\theta
^{^{\prime \prime }}\theta ^{^{\prime 2}}+\frac{1}{\tilde{T}^{4/3}}\sin
(2\theta )=0  \label{euler}
\end{equation}

\begin{figure}[tbp]
\includegraphics[width=3in,angle=0]{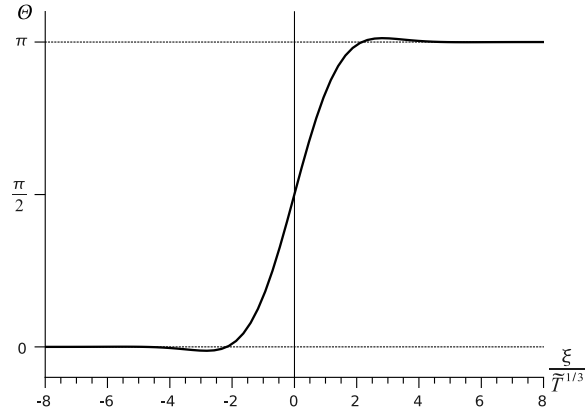}
\caption{Instanton solution of Eq.(\ref{euler}) with boundary
conditions: $\theta (-\infty )=0$, $\theta (\infty )=%
\pi$.}
\label{kink}
\end{figure}

We note that there are a few classical solutions of this Euler equation. Two
of them $\theta =0$ and $\theta =\pi $ correspond to trivial ferromagnetic
configurations. A systematic expansion around these saddle points is
equivalent to a purely perturbation expansion of the eigenvalues of the Schr%
\"{o}dinger equation (\ref{sch3d}). The corresponding contribution to the
partition function is $Z^{(0)}=\exp (-\lambda E_{0})$ where $E_{0}$ is given
by Eq.(\ref{ep}) and $\lambda $ is the scaled system length (\ref{lambda}).
The tunnel splitting is given by an instanton contribution to the functional
integral (\ref{Z3d}) \cite{book}. The classical solution of the Euler
equation corresponding to the instanton satisfies the boundary condition: $%
\theta (-\infty )=0$, $\theta (\infty )=\pi $ or $\theta (-\infty )=\pi $, $%
\theta (\infty )=0$. This solution of the Euler equation can be found
numerically. Actually, it coincides with the solution for the kink
excitation in the F-AF chain found by us in Ref.\cite{DK10}. The dependence $%
\theta (\xi )$ in this solution is shown in Fig.\ref{kink}. According to the
result of Ref.\cite{DK10}, the classical action corresponding to instanton
is $S_{0}=2/\tilde{T}$. The summation of the contributions to $Z$ using the
semiclassical approximation can be performed by a standard way \cite{book}.
As a result, the partition function is represented in a form
\begin{equation}
Z=Z^{(0)}+Z^{(2)}+Z^{(4)}+...  \label{Zsum}
\end{equation}%
where $Z^{(0)}$ is defined above, $Z^{(2)}$ is the instanton-antiinstanton
contribution (IA), $Z^{(4)}$ is the IAIA contribution and so on.

Using the usual approximation of the semiclassical method (in particular,
neglecting instanton-instanton interactions) we have
\begin{eqnarray}
Z^{(2)} &=&\frac{\lambda ^{2}e^{-2S_{0}}}{2}e^{-\lambda E_{0}}  \nonumber \\
Z^{(4)} &=&\frac{\lambda ^{4}e^{-4S_{0}}}{4!}e^{-\lambda E_{0}}
\end{eqnarray}
and so on (here $\lambda $ is the scaled system length (\ref{lambda})).

Summing (\ref{Zsum}) we arrive at
\begin{equation}
Z=e^{-\lambda E_{0}}\cosh (\lambda e^{-S_{0}})
\end{equation}

On the other hand, the partition function at $\lambda \to \infty $ can be
represented as
\begin{equation}
Z=e^{-\lambda E_{s}}+e^{-\lambda E_{a}}
\end{equation}
where $E_{s}$ and $E_{a}$ are the energies of the lowest states with even
and odd parity with respect to exchange $(r,\theta )\leftrightarrow (-r,\pi
-\theta )$. Then, $E_{s}=E_{0}-e^{-S_{0}}$ and $E_{a}=E_{0}+e^{-S_{0}}$. The
tunnel splitting is
\begin{equation}
\Delta E=E_{s}-E_{a}=2\exp (-2/\tilde{T})
\end{equation}

Then the susceptibility at $\tilde{T}\to 0$ to the exponential accuracy is
given by:
\begin{equation}
\widetilde{\chi }\sim \exp \left[ \frac{2s^{2}(\Delta -1)^{3/4}}{T}\right]
\label{chidd}
\end{equation}

The thermal gap in Eq.(\ref{chidd}) is the kink energy of the weakly
anisotropic classical F-AF chain. It is interesting to compare $\widetilde{%
\chi }$ (Eq.(\ref{chidd})) with the susceptibility of the quantum F-AF model
at $\alpha =\frac{1}{4}$ \cite{DK09}. The susceptibility for both models
shows the exponential dependence with the thermal gap proportional to $%
(\Delta -1)^{3/4}$. If we use Eq.(\ref{chidd}) for $s=\frac{1}{2}$ case we
find that the thermal gap is $\frac{1}{2}(\Delta -1)^{3/4}$ while in fact it
is $0.35(\Delta -1)^{3/4}$ \cite{DK09}, i.e. the numerical coefficients at $%
(\Delta -1)^{3/4}$ are slightly different.

\begin{figure}[tbp]
\includegraphics[width=3in,angle=0]{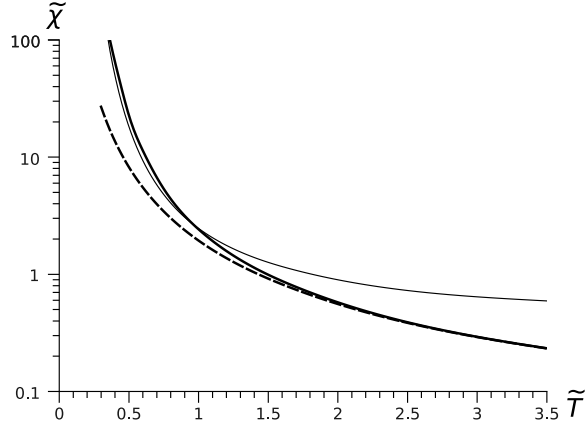}
\caption{Dependence of the normalized magnetic susceptibility $\widetilde{%
\chi}=(\Delta -1)\chi$ on the normalized temperature $\tilde{%
T}=T/s^{2}(\Delta -1)^{3/4}$ (thick solid line). Dashed and thin solid lines
are asymptotics of $\widetilde{\chi}$ at $\tilde{T}\to\infty$ and $%
\tilde{T}\to 0$ given by Eqs.(\ref{chis}) and (\ref{chidd}),
correspondingly.} \label{anisot}
\end{figure}

Eqs.(\ref{chis}) and (\ref{chidd}) give asymptotics of $\widetilde{\chi }(%
\tilde{T})$ for small and large values of $\tilde{T}$. In general case the
function $\widetilde{\chi}(\tilde{T})$ has been calculated numerically and
the dependence $\widetilde{\chi }(\tilde{T})$ is shown in Fig.\ref{anisot}
together with asymptotics of $\widetilde{\chi}$ for small and large $\tilde{T%
}$. As it can be seen from Fig.\ref{anisot} the dependence $\widetilde{\chi}$
on $\tilde{T}$ is characterized by two types of behavior: $\widetilde{\chi}$
is proportional to $\tilde{T}^{-4/3}$ at $\tilde{T}\gg 1$ and grows
exponentially at $\tilde{T}\to 0$. The crossover between two regimes occurs
at $\tilde{T}\sim 1$.

The calculation of the susceptibility of the anisotropic F-AF model can be
expanded to the case $\gamma >0$. We do not dwell on details of these
calculations. We notice only that $\widetilde{\chi }$ becomes the function
of $\tilde{T}$ and of a parameter $\mu =(4\alpha -1)/\sqrt{\Delta -1}$. $%
\widetilde{\chi }(\mu ,\tilde{T})$ as a function of $\tilde{T}$ has a
minimum for $\mu >\mu _{0}\simeq 1$ and is finite at $\tilde{T}\to 0$. The
normalized susceptibility $\widetilde{\chi }$ diverges at $\tilde{T}\to 0$
for $\mu <\mu _{0}$. The line $(4\alpha -1)=\mu _{0}\sqrt{\Delta -1}$ can be
identified with the boundary between the ferromagnetic and the helical phase.

Finally we give the results for the magnetization of the anisotropic F-AF
model at $\alpha =\frac{1}{4}$. According to Eqs.(\ref{chis}) and (\ref%
{chidd}) $m(g)$ at $g\to 0$ is
\begin{eqnarray}
m(g) &=&1.07g(1+0.81\delta ),\;\delta \ll 1  \nonumber \\
m(g) &\sim &g\exp (2\delta ^{3/4}),\;\delta \gg 1  \label{mg1}
\end{eqnarray}

The behavior of the magnetization in the high magnetic field limit is
obtained by analogy with that for the isotropic case (see Eq.(\ref{mtp})).
Then $m(g)$ at $g\gg 1$ is
\begin{equation}
m=1-\frac{1}{2(g+2\delta )^{3/4}}+O\left( (g+\delta )^{-3/2}\right)
\label{mg2}
\end{equation}

The magnetization curves $m(g)$ for several values of $\delta $ obtained by
the numerical solution of Eq.(\ref{sch3d}) is shown in Fig.\ref{delta}. Its
behavior in the limits $g\ll 1$ and $g\gg 1$ agrees with Eqs.(\ref{mg1}) and
(\ref{mg2}).

\section{Conclusions}

We have studied the low-temperature magnetic properties of the classical
anisotropic F-AF chain in the vicinity of the transition point from the
ferromagnetic to the helical ground state. This means that the frustration
parameter $\alpha =J_{2}/|J_{1}|$ is close to its critical value $\alpha=%
\frac{1}{4}$ and the anisotropy of the exchange interaction $(\Delta -1)$ is
weak. In the vicinity of the transition point the nearest spins in the
ground state are directed almost (or even exactly) parallel to each other.
Therefore, in the low-temperature limit when the thermal fluctuations are
weak, we can use the continuum approximation and represent the partition
function as a functional integral over the spin vector field. In the
obtained energy functional the model parameters $(\alpha -\frac{1}{4})$, $%
(\Delta -1)$ and the magnetic field $h$ are scaled by the temperature and
form three independent scaling parameters $\gamma $, $\delta $ and $g$
defined in Eq.(\ref{scpar}). This implies that we considered the scaling
limit when $\alpha \to \frac{1}{4}$, $\Delta \to 1$, $h\to 0$ and $T\to 0$,
but the values of the scaling parameters $\gamma $, $\delta $ and $g$ are
finite and govern the low-temperature thermodynamics of the F-AF model near
the transition point.

The derived functional integral for the partition function was treated as a
path integral of the quantum mechanics. The peculiarity of this path
integral is that the Lagrangian contains the second order derivative. To
handle with this problem we used the special Ostrogradski prescription,
which allowed us to obtain the quantum Hamiltonian corresponding to such
path integral in a special unusual form. Then the dependence of the lowest
eigenvalue of the Hamiltonian on the scaling parameters determines the
magnetization curves of the system. The eigenvalue problem has been solved
numerically and explicit expressions for the magnetization was obtained in
the limits of low and high magnetic fields.

It is known \cite{Sachdev} that the magnetization curve for the pure
ferromagnetic chain has a universal form when plotted against the
scaled magnetic field $g_{F}=s^{3}h/T^{2}$, and this curve is valid
for any value of spin $s$ including the classical limit $s\to \infty
$. We suppose that such universality remains for F-AF model at the
transition point against the scaling parameter $g=hs^{5/3}/T^{4/3}$.
If this is the case the obtained magnetization curves for the
classical model can be easily recalculated to the quantum spin case.
To validate this hypothesis one needs
to compare the obtained classical results with the magnetization of the $s=%
\frac{1}{2}$ F-AF model. Unfortunately, the exact thermodynamics of the
latter model is unknown. Nevertheless, there are two indirect arguments
supporting this conjecture. First is that the obtained critical exponent $%
\frac{4}{3}$ in the temperature dependence of susceptibility at the
transition point coincides with that obtained in the MSWT method.
The second argument is that three leading terms of the spin-wave
expansion of the magnetization of the quantum model coincide with
those for the classical model. Certainly these two facts do not
prove the proposed hypothesis and the question about its validity
remains open \cite{Sirker}. In this respect the numerical
calculations of the magnetization as a function of the magnetic
field at $T\to 0$ for $s=\frac{1}{2}$ and $s=1$ are very desirable.

Probably, the hypothesis of the universality of the function $m(g)$ (if any)
breaks down for $\alpha >\frac{1}{4}$ because the excitations above the
ground state are different in the quantum and in the classical F-AF chain.
Nevertheless, as was shown in Ref.\cite{DK11} some peculiarities of the
low-temperature behavior of the classical model at $\alpha >\frac{1}{4}$ is
qualitatively similar to that for the quantum $s=\frac{1}{2}$ chain. For
example, the temperature dependence of the zero-field susceptibility is in a
qualitative agreement with the numerical data for the quantum $s=\frac{1}{2}$
model and is in accord with the experimental data for the real edge-sharing
compounds.

We have studied the influence of the easy-axis anisotropy on the behavior of
the susceptibility at the transition point. It is shown that even weak
anisotropy essentially changes $\chi $. In the low-temperature limit the
susceptibility diverges exponentially in contrast with the isotropic case
where the divergence is of a power-like type. We note that such behavior of
the susceptibility takes place in the quantum $s=\frac{1}{2}$ F-AF chain
\cite{DK09} and the corresponding thermal gap has the same functional form
as the classical one. This fact confirms the close relation between the
low-temperature magnetic properties of the quantum and classical F-AF model
in the ferromagnetic part of the phase diagram.

\begin{acknowledgments}
We would like to thank S.-L.Drechsler and J.Sirker for valuable comments
related to this work.
\end{acknowledgments}

\end{document}